\documentclass[conference]{IEEEtran}
\usepackage{cite}

\ifCLASSINFOpdf
  \usepackage[pdftex]{graphicx}
  \graphicspath{{figs/}}
\else
\fi
\usepackage{amsmath}
\usepackage{algpseudocode}

\usepackage{array}
\newcolumntype{C}[1]{>{\centering\let\newline\\\arraybackslash\hspace{0pt}}m{#1}}

\ifCLASSOPTIONcompsoc
 \usepackage[caption=false,font=normalsize,labelfont=sf,textfont=sf]{subfig}
\else
 \usepackage[caption=false,font=footnotesize]{subfig}
\fi

\usepackage{stfloats}
\usepackage[squaren,binary,thinspace,thinqspace,textstyle]{SIunits}
\begin{document}

%
\title{Performance Analysis and Optimization of Sparse Matrix-Vector
  Multiplication on Modern Multi- and Many-Core Processors}

\author{\IEEEauthorblockN{Athena Elafrou\IEEEauthorrefmark{1},
Georgios Goumas\IEEEauthorrefmark{2}, and
Nectarios Koziris\IEEEauthorrefmark{3}}
\IEEEauthorblockA{School of Electrical and Computer Engineering\\
National Technical University of Athens}
\IEEEauthorblockA{Athens, Greece}
\IEEEauthorblockA{Email: 
\IEEEauthorrefmark{1}athena@cslab.ece.ntua.gr
\IEEEauthorrefmark{2}goumas@cslab.ece.ntua.gr
\IEEEauthorrefmark{3}nkoziris@cslab.ece.ntua.gr
}}

\maketitle

\begin{abstract}
  This paper presents a low-overhead optimizer for the ubiquitous
  sparse matrix-vector multiplication (SpMV) kernel. Architectural
  diversity among different processors together with structural
  diversity among different sparse matrices lead to bottleneck
  diversity. This justifies an SpMV optimizer that is both matrix- and
  architecture-adaptive through runtime specialization. To this
  direction, we present an approach that first identifies the
  performance bottlenecks of SpMV for a given sparse matrix on the
  target platform either through profiling or by matrix property
  inspection, and then selects suitable optimizations to tackle those
  bottlenecks. Our optimization pool is based on the widely used
  Compressed Sparse Row (CSR) sparse matrix storage format and has low
  preprocessing overheads, making our overall approach practical even
  in cases where fast decision making and optimization setup is
  required. We evaluate our optimizer on three x86-based computing
  platforms and demonstrate that it is able to distinguish and
  appropriately optimize SpMV for the majority of matrices in a
  representative test suite, leading to significant speedups over the
  CSR and Inspector-Executor CSR SpMV kernels available in the latest
  release of the Intel MKL library.
\end{abstract}


%
\IEEEpeerreviewmaketitle

\section{Introduction}
Sparse matrix-vector multiplication (SpMV) is a fundamental building
block of iterative methods for the solution of large sparse linear
systems and the approximation of eigenvalues of large sparse matrices
arising in applications from the scientific computing, machine
learning and graph analytics domains. It has been repeatedly reported
to achieve only a small fraction of the peak performance of current
computing systems due to a number of inherent performance limitations,
that arise from the algorithmic nature of the kernel, the employed
matrix storage format and the sparsity pattern of the matrix. In
particular, SpMV is characterized by a very low flop:byte ratio, since
the kernel performs $O(NNZ)$ operations on $O(N + NNZ)$ amount of
data\footnote{We define $NNZ$ as the number of nonzero elements and
  $N$ as the number of rows/columns in the matrix. For simplicity, we
  assume a square matrix.}, indirect memory references as a result of
storing the matrix in a compressed format and irregular memory
accesses to the right-hand side vector due to sparsity.

\begin{figure}[!t]
  \centering
  \includegraphics[width=0.48\textwidth]{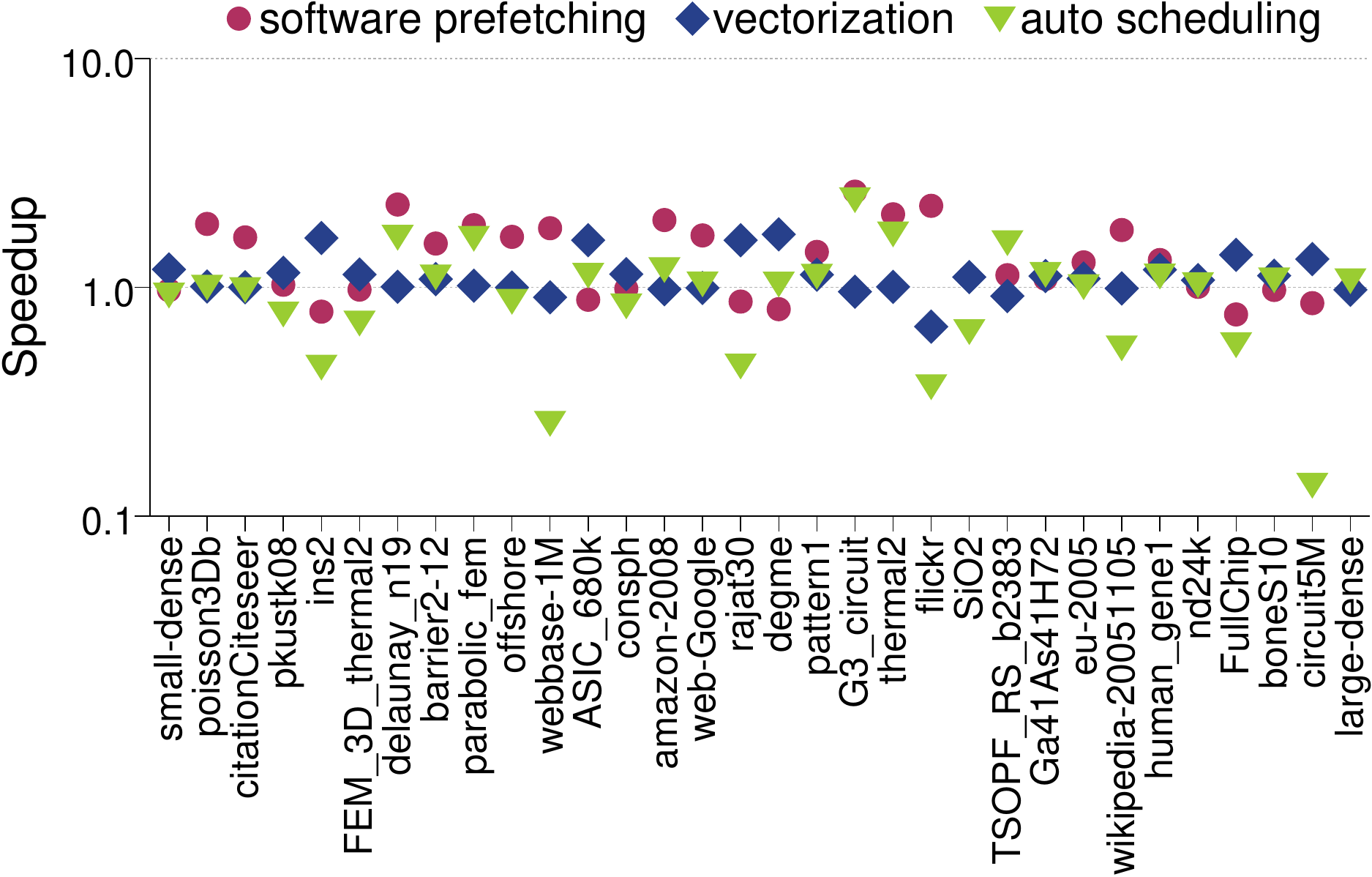}
  \caption{Speedup (slowdown) of different software optimizations
    applied to the CSR SpMV kernel on Intel Xeon Phi (codename Knights
    Corner).}
  \label{fig:intro}
  \vspace{-3mm}
\end{figure}

SpMV is typically a memory bandwidth bound kernel for the majority of
sparse matrices on multicore platforms. Its bandwidth utilization is
strongly dependent on the sparsity pattern of the matrix and the
underlying computing platform. Consequently, most optimization efforts
proposed in the literature over the past years have focused on
reducing traffic between caches and main memory, primarily by
compressing the memory footprint of the
matrix~\cite{toledo1997improving, pinar1999improving,
  pichel2004improving, vuduc2005fast, willcock2006accelerating,
  kourtis2008optimizing, williams2009optimization, bulucc2009parallel,
  belgin2009pattern, kourtis2011csx, bulucc2011reduced, 
  kreutzer2014unified}. With the advent of many-core architectures,
including GPGPUs and the Intel Xeon Phi processors, the performance
landscape has become more diverse. For instance, the larger number of
cores available in such architectures make SpMV performance more
sensitive to the workload distribution among threads. As the
performance of SpMV is becoming increasingly dependent on both the
input problem and the underlying computing platform, there is no
one-size-fits-all solution to attain high performance. Thus, we
advocate the significance of matrix- and architecture-aware runtime
specialization. The value of cherry-picking optimizations for SpMV is
further emphasized by the fact that blindly applying optimizations can
actually hinder performance. As an example, Fig.~\ref{fig:intro} shows
the effect of different optimizations on an Intel Xeon Phi
processor. Even though each optimization achieves significant gains
for some matrices, it may result in nonnegligible slowdowns for
others.
Optimizing SpMV usually involves a preprocessing step to analyze the
sparse matrix structure and may include format conversion, parameter
tuning, etc. While the overhead of this step may be amortized for
applications that reuse the same matrix over many invocations of SpMV,
it can outweigh any performance benefit when a smaller number of
iterations is required for solver convergence, which is often the case
in preconditioned solvers~\cite{li2013gpu}, or when the structure of
the matrix changes frequently, e.g. in graph applications. This is why
many recent efforts have focused on designing more lightweight
optimizations~\cite{Greathouse:2014:ESM:2683593.2683678,
  Liu:2015:CES:2751205.2751209}.

In this paper, we propose an adaptive SpMV optimizer that seeks to
provide \emph{performance stability}, i.e., improved performance for
all sparse matrices, and \emph{low overhead} in order to be beneficial
to problems that require a small number of iterations to converge. We
especially focus on designing an optimizer that is more lightweight
than a trivial one that runs all optimizations and selects the best
for the target application.

In order to provide performance stability, our optimizer relies on
SpMV bottleneck analysis of the input matrix on the target
platform. We formulate bottleneck detection as a classification
problem (Section~\ref{sec:methodology:classes}) and first develop a
profile-guided classifier
(Section~\ref{sec:methodology:classifiers:profiling}), that relies on
micro-benchmarks to classify the matrix. As the online profiling phase
has a nonnegligible cost, we go one step beyond, and propose a
classifier that relies only on matrix properties that can be cheaply
computed (Section~\ref{sec:methodology:classifiers:feature}). This
feature-guided classifier is pre-trained during an offline stage with
the use of machine learning techniques and performs feature extraction
on-the-fly. The runtime overhead of this classifier is very low,
making it extremely lightweight (Section~\ref{sec:eval:overhead}).

Once the performance bottlenecks of a matrix have been detected, we
apply suitable optimizations to tackle them. We employ a simple and
easy-to-implement set of optimizations (Section~\ref{sec:pool}), based
on the general-purpose Compressed Sparse Row (CSR) sparse matrix
storage format. Our approach, however, is compatible with a plethora
of other optimizations found in the literature. Again here, the idea
is to resort to optimizations that require short preprocessing times,
further contributing to our objective for a lightweight scheme.

 
We evaluate our optimizer using both classifiers on two generations of
Intel's Xeon Phi many-core processor and an Intel Broadwell
processor. The performance analysis in Section~\ref{sec:methodology}
demonstrates the diversity of SpMV bottlenecks among different
matrices, while our experimental results in
Section~\ref{sec:evaluation} show the diversity of SpMV bottlenecks
among different computing platforms and the effectiveness of the
proposed approach in distinguishing and appropriately optimizing the
majority of matrices, leading to significant speedups over the CSR and
Inspector-Executor CSR SpMV implementations available in the latest
Intel MKL library, which is highly optimized for x86-based platforms.

\section{Background}
\label{sec:background}
In order to avoid the extra computation and storage overheads imposed
by the large majority of zero elements found in a sparse matrix, it
has been the norm to store the nonzero elements contiguously in memory
and employ auxiliary data structures for their proper traversal. The
most widely-used general-purpose sparse matrix storage format, namely
the \textit{Compressed Sparse Row} (CSR)
format~\cite{saad1992numerical}, uses a row pointer array to index the
start of each row within the array of nonzero elements, and a column
index array to store the column of each nonzero element. The $y =
A\cdot x$ SpMV kernel using this format is given in
Fig.~\ref{algo:spmv}.

\begin{figure}[!t]
  \footnotesize
  \begin{algorithmic}[1]
    \Procedure{Spmv}{$A$::in, $x$::in, $y$::out}
    \Statex{$A$: matrix in CSR format}
    \Statex{$x$: input vector}
    \Statex{$y$: output vector}
    \For{$i \gets 1,N$}
    \For{$j \gets A.rowptr[i],A.rowptr[i+1]$}
    \State{$y[i] \gets A.val[j]\cdot x[A.colind[j]]$}
    \EndFor
    \EndFor
    \EndProcedure
  \end{algorithmic}
  \caption{SpMV implementation using the CSR format.}
  \label{algo:spmv}
  \vspace{-7mm}
\end{figure}

SpMV is characterized by a \emph{low operational intensity}. It
performs $O(NNZ)$ floating-point operations on $O(NNZ + N)$ amount of
data, leading to a flop:byte ratio of less than 1. According to the
Roofline performance model~\cite{williams2009roofline}, computational
kernels with low operational intensities tend to be memory
bound. Furthermore, the auxiliary indexing structures required to
access the nonzero elements, namely \texttt{rowptr} and
\texttt{colind} in case of CSR, introduce additional load operations,
traffic for the memory subsystem, and cache
interference~\cite{pinar1999improving}. Access to the input vector $x$
is irregular and depends on the sparsity pattern of the matrix. This
fact complicates the process of exploiting any spatial or temporal
reuse. Many sparse matrices contain a large number of short rows that
may degrade performance on par with the small trip count of the inner
loop~\cite{mellor2004optimizing}. Finally, many sparse matrices are
characterized by rows with highly uneven lengths or regions with
different sparsity patterns, which may result in workload imbalance~\cite{tang2015optimizing}.
%
\section{Optimization Tuning Methodology}
\label{sec:methodology}
Optimization selection can be solved in various ways. For instance,
one could simply take an empirical approach: measure how different
optimizations work for a particular matrix on the target machine and
then apply the most efficient optimization on future runs of the same
matrix. This is the most accurate method, however, it incurs a
substantial overhead to sweep over all candidate optimizations, some
of them requiring significant preprocessing time including format
conversion and autotuning parameters. In order to reduce this
overhead, one could employ a ``blind'' machine learning approach. By
defining optimizations as classes, one could train a classifier with
machine learning techniques to select a proper optimization. However,
in this case, the classifier would need to be rebuilt whenever an
optimization is added or replaced and, furthermore, no intuition would
be provided into when and why an optimization works.

We propose a solution that is lightweight, intuitive and enables a
modular framework design. We formulate optimization selection as a
multiclass, multilabel classification problem, where classes represent
performance bottlenecks. For every class, we implement an optimization
that is designed to address the corresponding bottleneck. Given an
input matrix, its bottlenecks are identified through classification
and the corresponding optimizations are jointly applied. Since most
optimizations proposed in the literature are designed with a specific
bottleneck in mind, by detecting bottlenecks instead of optimizations,
our classifier explains why an optimization actually improved
performance for a particular matrix. Furthermore, by decoupling
bottleneck identification from the application of optimizations, one
can build a classifier once and optimizations can be henceforth added
or replaced in a plug-and-play fashion.




\subsection{Formulation as a Classification Problem}
\label{sec:methodology:classes}
Based on the analysis presented in Section~\ref{sec:background} and
the extensive research that has been realized over the past few
decades in SpMV performance analysis and optimization, we define the
following classes, each representing a potential SpMV performance
bottleneck:

\begin{itemize}[\IEEEsetlabelwidth{ML}]
\item[\emph{MB}] This class includes matrices that achieve memory
  bandwidth utilization values close to the peak (which may differ
  depending on the working set size) and are, therefore, \emph{Memory
    Bandwidth} bound. Such matrices usually exhibit a regular sparsity
  structure.
\item[\emph{ML}] This class refers to matrices that are \emph{Memory
  Latency} bound. Such matrices exhibit poor locality in the accesses
  to the right-hand side vector due to a highly irregular sparsity
  pattern, which cannot be detected by hardware prefetching mechanisms
  available in current architectures.
\item[\emph{IMB}] This class includes matrices with highly uneven row
  lengths or regions with completely different sparsity patterns that
  exhibit thread \emph{IMBalance} due respectively to an uneven
  workload distribution or computational unevenness.
\item[\emph{CMP}] This class includes matrices that present
  \emph{CoMPutational} bottlenecks. Such matrices either fit in the
  system's cache hierarchy and have an operational intensity closer to
  the corresponding dense kernel, pushing them closer to the ridge
  point of the Roofline model~\cite{williams2009roofline}, or have the
  majority of nonzero elements concentrated in a few dense rows,
  e.g. power law matrices.
\end{itemize}

The above classes are quite generic and cover a wide variety of
matrices with different structural characteristics.

\subsection{Per-Class Performance Bounds}
\label{sec:methodology:bounds}
In this section, we perform a thorough \emph{bound and bottleneck
  analysis} for the SpMV kernel. Inspired by the ``bounding
techniques'' described in~\cite{lazowska1984quantitative}, for every
class defined in Section~\ref{sec:methodology:classes}, we derive an
upper bound on performance that represents the maximum performance
that can be achieved by completely eliminating the corresponding
bottleneck. The idea here is that by comparing the actual performance
with the per-class upper bounds, we are able to estimate which
bottlenecks are important to address and what is the potential
performance gain. In the following analysis, $P$ stands for
performance, $M$ for memory traffic, $B$ for bandwidth, $S$ for size
and $t$ for execution time.

\begin{itemize}[\IEEEsetlabelwidth{ML}]
\item[$P_{MB}$] For matrices of this class the available memory
  bandwidth is saturated. Thus, an upper performance bound can be
  derived by assuming maximum sustainable main memory bandwidth as
  \begin{equation*}
    P_{MB} = \frac{2\cdot NNZ}{\frac{M_{A_{format},min} + M_{xy,min}}{B_{max}}}
  \end{equation*}
  \noindent
  where $B_{max}$ is the maximum sustainable memory bandwidth of the
  system\footnote{We adjust the bandwidth upwards for matrices that
    fit in the system's cache hierarchy.}, while $M_{A_{format},min}$
  and $M_{xy,min}$ are the minimum memory traffic that can be
  generated by the matrix $A$ stored in \emph{format} and the vectors
  $x,y$ respectively. Note that compulsory misses set the minimum
  memory traffic and, hence, $M_{A_{format},min} = S_{format}$ and
  $M_{xy,min} = S_{x} + S_{y}$.
\item[$P_{ML}$] For this class, we can derive an upper bound by
  running a modified SpMV kernel where irregular accesses to the
  right-hand side vector $x$ are converted to regular accesses. In
  case of the CSR format, we can achieve this by setting all entries
  of the \texttt{colind} array to the row index of the corresponding
  element.
\item[$P_{IMB}$] An upper bound for this class can be estimated using
  the median execution time $t_{median}$ among all threads as
  \begin{equation*}
    P_{IMB} = \frac{2\cdot NNZ}{t_{median}}
  \end{equation*}
  We use the median instead of the mean, as we require reduced
  importance to be attached to outliers.
\item[$P_{CMP}$] For matrices of this class, we can define a very
  loose upper bound by running a modified SpMV kernel where we
  completely eliminate indirect memory references, resulting in
  unit-stride accesses only. In case of the CSR format, we no longer
  use \texttt{colind} to index vector $x$, but always access $x[i]$,
  where $i$ is the current row index.
\end{itemize}

A generic upper bound on SpMV performance, independent of the format
used to store the sparse matrix, can be defined as follows:
\begin{equation*}
  P_{peak} = \frac{2\cdot NNZ}{\frac{M_{A,min} + M_{xy,min}}{B_{max}}}
\end{equation*}
\noindent
where $B_{max}$ and $M_{xy,min}$ are defined as before. In order to
estimate $M_{A,min}$ we assume we can only compress the indexing
information of a sparse matrix (not the values). In this case, a
minimum can be determined by completely eliminating the indexing
structures, i.e., $M_{A,min} = S_{values}$.

We apply the bound and bottleneck analysis on an Intel Xeon Phi
co-processor using the CSR format as our baseline SpMV
implementation. You may refer to Section~\ref{sec:evaluation:setup}
for details on the experimental setup. The performance of the baseline
kernel is denoted by $P_{CSR}$. Among the previously defined
performance bounds, only $P_{ML}$ and $P_{CMP}$ require a
micro-benchmark to be run on-the-fly. $P_{peak}$ and $P_{MB}$ only
require $B_{max}$ to be measured on the target system, while $P_{IMB}$
can be deduced from the baseline run. Fig.~\ref{fig:bounds} shows the
performance of the baseline along with the per-class and generic upper
bounds for a representative set of matrices from the University of
Florida Sparse Matrix Collection~\cite{davis2011university}. The
relation of each per-class upper bound to the baseline differs among
matrices, indicating a potential diversity in the bottlenecks of each
matrix.

\subsection{Profile-Guided Classifier}
\label{sec:methodology:classifiers:profiling}
By comparing the per-class upper bounds to the baseline performance
presented in the previous section, we are able to derive heuristics
that can be used to determine the class(es) of each matrix. Examining
Fig.~\ref{fig:bounds} in more detail, we distinguish the following
cases:

\begin{figure}[!t]
  \centering
  \includegraphics[width=0.48\textwidth]{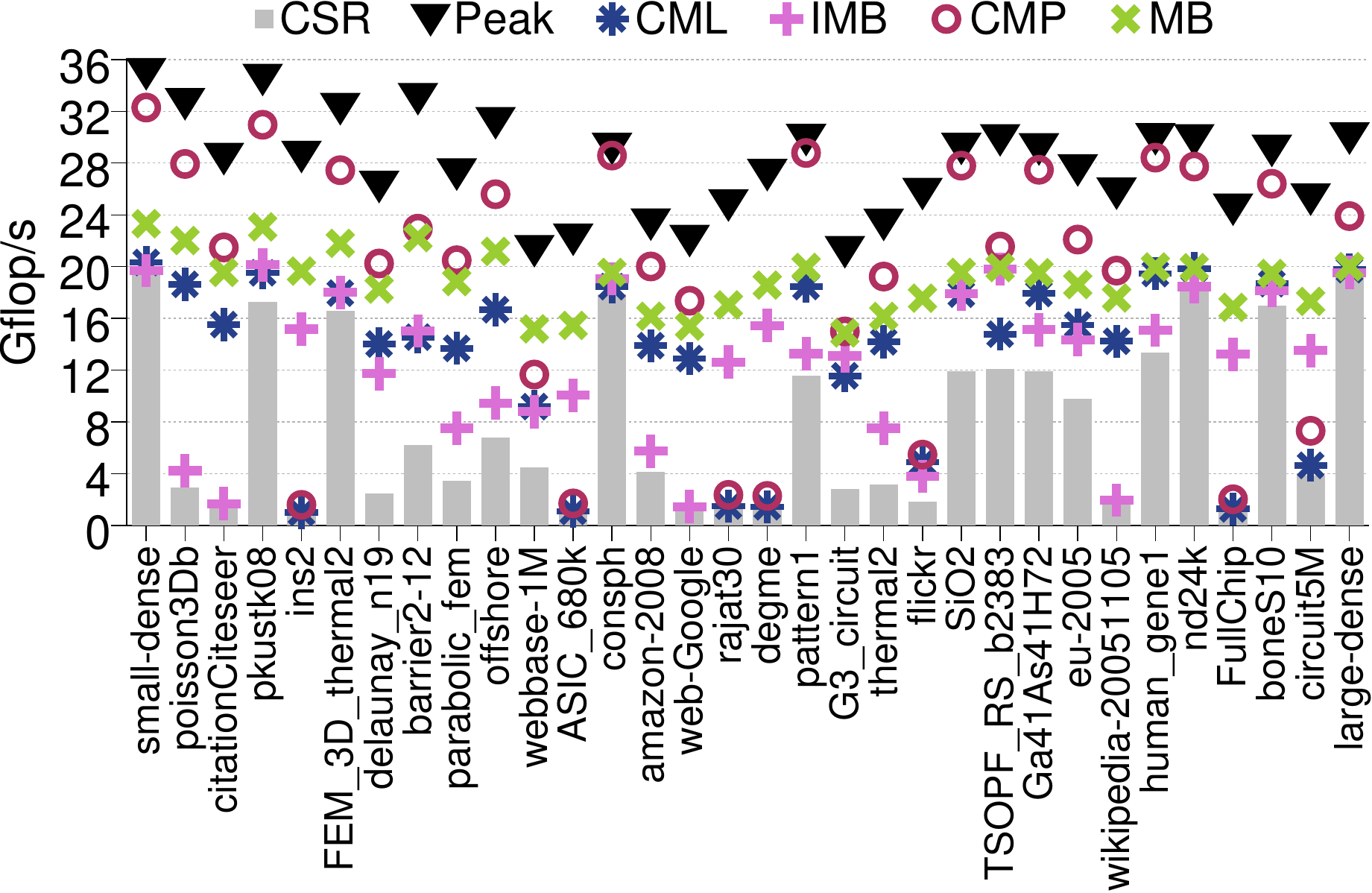}
  \caption{SpMV performance using the CSR format and per-class upper
    bounds on Intel Xeon Phi (codename Knights Corner).}
  \label{fig:bounds}
  \vspace{-3mm}
\end{figure}

\begin{itemize}
\item $P_{CSR}\approx P_{ML}$ (\emph{consph, rajat30, circuit5M},
  etc.)  Performance did not improve by eliminating irregular memory
  accesses, hence SpMV does not suffer from excessive cache
  misses. Note that these matrices incurred a slowdown with software
  prefetching in Fig.~\ref{fig:intro}.
\item $P_{CSR}\approx P_{IMB}$ (\emph{citationCiteseer, web-Google,
  nd24k}, etc.) This implies a balanced execution time among all
  threads.
\item $P_{CSR}\approx P_{MB}$ (\emph{consph, nd24k, boneS10}, etc.) In
  this case, either SpMV has saturated the available main memory
  bandwidth and, thus, the matrix is a candidate of the MB class.
\item $P_{ML} \gg P_{CSR}$ and/or $P_{IMB} \gg P_{CSR}$
  (\emph{poisson3Db, parabolic\_fem, offshore}) In this case CSR SpMV
  benefited from eliminating irregularity and/or balancing thread
  execution time, indicating it may belong to the ML and/or IMB
  classes.
\item $P_{CMP} \ll P_{MB}$ (\emph{ASIC\_680k, rajat30, degme}) This
  suggests that SpMV using the CSR format might be limited by
  computational bottlenecks, based on the following analysis:
  \begin{IEEEeqnarray}{rCl}
    P_{CMP} < P_{MB} \Leftrightarrow \frac{P_{CMP}}{P_{MB}} < 1 &&
    \Leftrightarrow \nonumber \\ \frac{\frac{2\cdot
        NNZ}{\frac{M_{CMP}}{B_{CMP}}}}{\frac{2\cdot
        NNZ}{\frac{M_{A_{CSR},min} + M_{xy,min}}{B_{max}}}} < 1 &&
    \Leftrightarrow \nonumber \\ \frac{M_{A_{CSR},min} +
      M_{xy,min}}{M_{A_{CSR},CMP} + M_{xy,CMP}}\cdot
    \frac{B_{CMP}}{B_{max}} < 1
    \label{eq:1}
  \end{IEEEeqnarray}
  \noindent
  Since the memory footprint of the matrix in the computation of
  $P_{CMP}$ is smaller (we no longer use the \texttt{colind} array)
  and we have eliminated irregularity as well, we can assume that
  $M_{A_{CSR},CMP} < M_{A_{CSR},min}$ and $M_{xy,CMP} <
  M_{xy,min}$. Thus, in order for~(\ref{eq:1}) to hold, $B_{CMP} <
  B_{max}$. Thus, the matrix is not memory bound (neither latency nor
  bandwidth). If additionally $P_{IMB} \gg P_{CSR}$ the execution is
  imbalanced and the threads causing the imbalance are
  compute-limited. This condition holds for matrices that have the
  majority of nonzero elements concentrated in a few dense rows, e.g.,
  \emph{ASIC\_680k, rajat30, degme, FullChip} and
  \emph{circuit5M}. Note that these matrices improve with
  vectorization in Fig.~\ref{fig:intro}. If $P_{CMP} \gg P_{ML}$, the
  matrix may be suffering from loop overheads in the presence of very
  short rows, e.g., \emph{webbase-1M, flickr, circuit5M}.
\item $P_{CMP} \gg P_{PEAK}$ Based on a theoretical analysis similar
  to the previous one, we conclude that this condition may occur when
  SpMV's working set fits in the system's cache hierarchy and the
  system's bandwidth vastly improves for such working sets. This case
  does not appear in our matrix suite on KNC, it does however on other
  platforms.
\end{itemize}

Our analysis clearly exposes a diversity in performance bottlenecks on
Intel Xeon Phi. Many matrices suffer from thread imbalance and
excessive cache misses, while a subset of those seem to have competing
bottlenecks, e.g.,\ \emph{parabolic\_fem, offshore, webbase-1M},
etc. This diversity justifies an SpMV optimizer that is
matrix-adaptive. Towards this direction, we design a rule-based
classification algorithm based on the previous observations. Since we
collect data for the performance bounds during an online profiling
phase, we will henceforth refer to this classifier as the
\textit{profile-guided} classifier.

The algorithm is depicted in Fig.~\ref{algo:classifier}. The
classification is performed by comparing per-class performance bounds
and by evaluating their impact on the baseline kernel. The values of
$T_{ML}$ and $T_{IMB}$, which form the hyperparameters of our
classifier, have been tuned using grid search, which simply performs
an exhaustive search through the specified hyperparameter space for a
combination of values that maximizes some performance metric. We
choose to maximize the average performance gain of the corresponding
optimizations on a large set of matrices (see
Section~\ref{sec:pool}). Notice that it is possible for a matrix not
to be classified. This reflects cases for which it is not worth
applying any of the optimizations in our pool. Also note that this
classifier implicitly deduces SpMV performance characteristics on the
underlying computing platform through hyperparameter tuning and
profiling, making it architecture-adaptive.

\begin{figure}[t]
  \footnotesize
  \begin{algorithmic}[1]\raggedright
    \Procedure{Classify}{$P_{CSR}$, $P_{MB}$, $P_{ML}$, $P_{IMB}$, $P_{CMP}$, $P_{peak}$}
    \State{$class \gets \O$}
    \If {$(\frac{P_{IMB}}{P_{CSR}} > T_{IMB})$}
    \State{$class \gets class \cup \left\{IMB\right\}$}
    \EndIf
    \If {$(\frac{P_{ML}}{P_{CSR}} > T_{ML})$}
    \State{$class \gets class \cup \left\{ML\right\}$}
    \EndIf
    \If {$(P_{CSR}\approx P_{MB}$ \textbf{and} $P_{MB} < P_{CMP} < P_{peak})$}
    \State{$class \gets class \cup \left\{MB\right\}$}
    \EndIf
    \If {$(P_{MB} > P_{CMP}$ \textbf{or} $P_{CMP} > P_{peak})$}
    \State{$class \gets class \cup \left\{CMP\right\}$}
    \EndIf
    \State \textbf{return} $class$
    \EndProcedure
  \end{algorithmic}
  \caption{Profile-guided classifier. Parameters $T_{ML}=1.25$ and
    $T_{IMB}=1.24$ were optimized through exhaustive grid search.}
  \label{algo:classifier}
  \vspace{-3mm}
\end{figure}

\subsection{Feature-Guided Classifier}
\label{sec:methodology:classifiers:feature}
We go one step further and design a classifier that relies on
structural features of the sparse matrix to perform the
classification. The intuition behind this approach is that one can
devise features that represent comprehensive characteristics of a
sparse matrix that are closely related to SpMV performance. For
example, the number of nonzero elements per row can reveal workload
imbalance during SpMV execution. In particular, if a matrix contains
very uneven row lengths and a row-partitioning scheme is being
employed for workload distribution, then threads that are assigned
long rows will take longer to execute resulting in thread imbalance.
We leverage supervised learning techniques to build this
classifier. The advantage of this approach is that, given a set of
labeled matrices, the classification rules can be automatically
deduced. We will henceforth refer to this classifier as the
\textit{feature-guided} classifier.


We experiment with a Decision Tree classifier and adjust it to perform
multilabel classification in order to detect all bottlenecks. We also
add a dummy class to reflect matrices not worth optimizing with any of
our optimizations (see Section~\ref{sec:pool}). Learning is performed
using an optimized version of the CART algorithm, which has a runtime
cost of $O(N_{features}\cdot N_{samples}\cdot \log N_{samples})$ for
the construction of the tree, where $N_{features}$ is the number of
features used and $N_{samples}$ is the number of samples used to build
the classifier. The query time for this classifier is $O(\log
N_{samples})$. We generate the classifier using the
\emph{scikit-learn} machine learning toolkit~\cite{scikit-learn}.

\subsubsection{Feature Extraction}
This classifier uses real-valued features to perform the
classification. Table~\ref{table:features} shows all the features we
experimented with, along with the time complexity of their extraction
from the input matrix. $N$ denotes the number of rows in the matrix
and $NNZ$ the number of nonzero elements. We define $nnz_{i}$ as the
number of nonzero elements of row $i$, $bw_{i}$ the column distance
between the first and last nonzero element of row $i$, $scatter_{i} =
\frac{nnz_{i}}{bw_{i}}$, $clustering_{i} = \frac{ngroups_i}{nnz_{i}}$,
where $ngroups_i$ is the number of groups formed by consecutive
elements in row $i$ and, finally, $misses_i$ is the number of nonzero
elements in row $i$ that can generate cache misses. We naively say
that an element will generate a cache miss when its distance from the
previous element in the same row exceeds the number of elements that
fit in a cache line of the system.

\begin{table}
  \renewcommand{\arraystretch}{1.5}
  \caption{Sparse matrix features used for classification}
  \label{table:features}
  \centering
  \begin{tabular}{|p{1.6cm}|c|c|}
    \hline
    \textbf{Feature} & \textbf{Definition} & \textbf{Complexity} \\
    \hline
    \centering$size$ & 0:exceeds or 1:fits in LLC & $\Theta(1)$ \\
    \centering$density$ & $\frac{NNZ}{N^{2}}$ & $\Theta(1)$ \\
    \centering$nnz_{\text{min}}$ & $\min\{nnz_{1},\dotsc,nnz_{N}\}$ & $\Theta(N)$ \\
    \centering$nnz_{\text{max}}$ & $\max\{nnz_{1},\dotsc,nnz_{N}\}$ & $\Theta(N)$ \\
    \centering$nnz_{\text{avg}}$ & $\frac{1}{N}\sum_{i=1}^{N}nnz_{i}$ & $\Theta(N)$ \\
    \centering$nnz_{\text{sd}}$ & $\sqrt{\frac{1}{N}\sum_{i=1}^{N} (nnz_{i}-nnz_{\text{avg}})^2}$ & $\Theta(2N)$ \\
    \centering$bw_{\text{min}}$ & $\min\{bw_{1},\dotsc,bw_{N}\}$ & $\Theta(N)$ \\
    \centering$bw_{\text{max}}$ & $\max\{bw_{1},\dotsc,bw_{N}\}$ & $\Theta(N)$ \\
    \centering$bw_{\text{avg}}$ & $\frac{1}{N}\sum_{i=1}^{N}bw_{i}$ & $\Theta(N)$ \\
    \centering$bw_{\text{sd}}$ & $\sqrt{\frac{1}{N}\sum_{i=1}^{N} (bw_{i}-bw_{avg})^2}$ & $\Theta(2N)$ \\
    \centering$scatter_{\text{avg}}$ & $\frac{1}{N}\sum_{i=1}^{N}scatter_{i}$ & $\Theta(N)$ \\
    \centering$scatter_{\text{sd}}$ & $\sqrt{\frac{1}{N}\sum_{i=1}^{N} (scatter_{i}-scatter_{avg})^2}$ & $\Theta(2N)$ \\
    \centering$clustering_{avg}$ & $\frac{1}{N}\sum_{i=1}^{N}clustering_{i}$ & $\Theta(NNZ)$ \\
    \centering$misses_{avg}$ & $\frac{1}{N}\sum_{i=1}^{N}misses_{i}$ & $\Theta(NNZ)$ \\
    \hline
  \end{tabular}
  \vspace{-3mm}
\end{table}

\subsubsection{Training Data Selection}
We train our classifier with a data set consisting of sparse matrices
from the University of Florida Sparse Matrix
Collection~\cite{davis2011university}. We have selected a matrix suite
consisting of 210 matrices from a wide variety of application domains,
to avoid being biased towards a specific sparsity pattern.

\subsubsection{Training Data Labeling}
Since the classes of a matrix cannot be determined in a
straightforward manner, we use our profile-guided classifier for this
purpose. An issue that arises by this choice is that the validity of
the labels and, consequently, the accuracy of the trained classifier,
depends on the behavior of the profile-guided classifier.

\subsection{Optimization Pool}
\label{sec:pool}
Table~\ref{table:optimizations} gives a mapping of classes to
optimizations. We have incorporated optimizations that require minimal
preprocessing time, further contributing to our objective for a
lightweight scheme. Note that in case multiple bottlenecks are
detected, we jointly apply the corresponding optimizations.

For matrices that are memory bandwidth limited, we employ a rather
simple compression scheme, just for illustrative purposes. We use
delta indexing on the column indices of the nonzero elements of the
matrix, a technique which was originally applied to SpMV by Pooch and
Nieder~\cite{pooch1973survey}. We use 8- or 16-bit deltas wherever
possible, but never both, in order to limit the branching overhead
during SpMV computation. We employ vectorization as well, which helps
attain a higher bandiwdth utilization value on modern processors with
wide SIMD units.

For the ML class of matrices, we employ software prefetching, since
the source of excessive cache misses in SpMV is a result of indirect
memory addressing to the right-hand side vector, which cannot be
efficiently tackled by current hardware prefetching mechanisms. A
single prefetch instruction was inserted in the inner loop of SpMV,
with a fixed prefetch distance equal to the number of elements that
fit in a single cache line of the hardware platform. Data are
prefetched into the L1 cache.

\begin{figure}[t]
  \centering
  \includegraphics[scale=0.7]{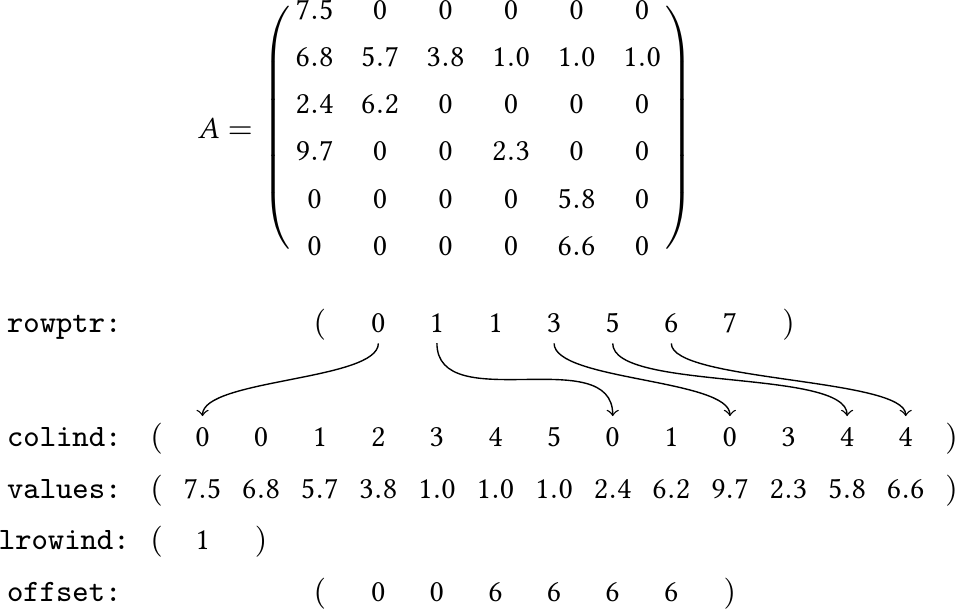}
  \caption{Matrix decomposition optimization for matrices with very long rows.}
  \label{fig:decomp}
  \vspace{-3mm}
\end{figure}

For matrices that suffer from thread imbalance, we include multiple
optimizations, as thread imbalance can be a result of either workload
imbalance (number of nonzero elements per thread) or computational
unevenness. We select the final optimization based on structural
features of the matrix. The first subcategory includes matrices with
highly uneven row lengths that we detect by comparing the $nnz_{max}$
and $nnz_{avg}$ features from Table~\ref{table:features}. For these
matrices, we basically decompose the matrix into two parts stored in a
modified CSR format, depicted in Fig.~\ref{fig:decomp}. SpMV is
performed in two steps, described in
Fig.~\ref{algo:spmv-decomp}. First, SpMV is performed as usual by
skipping the long rows, and, then long rows are computed with a
different assignment of computation to threads. Every row is computed
by all threads and a reduction of partial results follows. For the
second subcategory, which we detect with the $bw_{sd}$ feature, we
employ the \textit{auto} scheduling policy available in the OpenMP
runtime system~\cite{dagum1998openmp}. When the \textit{auto} schedule
is specified, the decision regarding scheduling is delegated to the
compiler, which has the freedom to choose any possible mapping of
iterations (rows in this case) to threads.

Finally, for matrices that present computational bottlenecks, we
combine inner loop unrolling with vectorization.

\begin{table}[t]
  \caption{Mapping of matrix classes to optimizations}
  \label{table:optimizations}
  \centering
  \begin{tabular}{|c|C{7.0cm}|}
    \hline
    \textbf{Class} & \textbf{Optimization} \\
    \hline
    MB & column index compression through delta encoding~\cite{pooch1973survey} + vectorization\\
    ML & software prefetching on vector x \\
    IMB & \vtop{\hbox{\strut matrix decomposition, \textit{auto} scheduling (OpenMP)}} \\
    CMP & inner loop unrolling + vectorization \\
    \hline
  \end{tabular}
\end{table}

\begin{figure}[t]
  \footnotesize
  \begin{algorithmic}[1]\raggedright
    \Procedure{Spmv}{$A$::in, $x$::in, $y$::out}
    \For{$i \gets 1,N$}
    \For{$j \gets A.rowptr[i], A.rowptr[i+1]$}
    \State{$off_{i} \gets A.offset[i]$}
    \State{$y[i] \gets A.val[j+off_{i}]\cdot x[A.colind[j+off_{i}]]$}
    \EndFor
    \EndFor
    \For{$i \gets 1,N'$}\Comment{$N'$: number of long lines}
    \State{$row \gets lrowind[i]$}
    \State{$off_{i} \gets A.offset[row]$}
    \State{$off_{i+1} \gets A.offset[row+1]$}
    \For{$j \gets A.rowptr[row]+off_{i},$ \\ $\qquad\qquad\qquad\;\: A.rowptr[row+1]+off_{i+1}$}
    \State{$y[row] \gets A.val[j + off_{i}]\cdot$\\$\qquad\qquad\qquad\qquad\quad x[A.colind[j+off_{i+1}]]$}
    \EndFor
    \EndFor
    \EndProcedure
  \end{algorithmic}
  \caption{Decomposed SpMV implementation for matrices with highly uneven row
    lengths.}
  \label{algo:spmv-decomp}
  \vspace{-6mm}
\end{figure}

\section{Experimental Evaluation}
\label{sec:evaluation}
Our experimental evaluation focuses on three aspects: matrix
classification accuracy, the performance benefit of applying our
methodology for optimizing SpMV, as well as the runtime cost of the
optimization workflow.

\subsection{Experimental Setup and Methodology}
\label{sec:evaluation:setup}
Our experiments were performed on a first generation Intel Xeon Phi
co-processor (KNC), a second generation Intel Xeon Phi bootable
processor (KNL) configured in the \emph{Flat mode}, in which the
high-bandwidth MCDRAM memory (HBM) is directly addressable, and an
Intel Broadwell processor (Broadwell). On KNL, we allocate the entire
application on HBM using the \texttt{numactl}
tool. Table~\ref{table:hw} lists the technical specifications of each
platform. We disabled the turbo boost technology to avoid fluctuations
in the measurements.

We used the OpenMP parallel programming interface and compiled our
software using Intel ICC-17.0.1 and the Intel OpenMP library. The
benchmarks used for the profile-guided classifier as well as the
baseline SpMV implementation use the CSR format and are compiled with
the \texttt{-O3 -qopt-prefetch=0 -qopt-threads-per-core=4} flags. To
simulate the typical SpMV computation in scientific applications, we
used double precision floating-point values for the nonzero
elements. Unless stated otherwise, the baseline and optimized
implementations employ a static one-dimensional row partitioning
scheme, where each partition has approximately equal number of nonzero
elements and is assigned to a single thread. For comparison purposes
we used Intel MKL 2017.0. Finally, we enforced the thread affinity
policy by setting \texttt{OMP\_PLACES=threads} and
\texttt{OMP\_PROC\_BIND=close} using all the cores in each system.

Whenever we report performance rates for SpMV, e.g. Gflop/s, they have
been summarized over 5 benchmark runs using the harmonic mean. The
rate of each individual run is the rate of the arithmetic means of the
absolute counts (floating-point operations and seconds) of 128 SpMV
operations, that is, we perform warm cache measurements.

\begin{table}[]
  \scriptsize
  \renewcommand{\arraystretch}{1.1}
  \caption{Technical characteristics of the experimental platforms}
  \label{table:hw}
  \centering
  \begin{tabular}{|C{2cm}|C{1.7cm}|C{1.7cm}|C{1.5cm}|}
    \hline
    \textbf{Codename} & \textbf{Knights Corner (KNC)} & \textbf{Knights Landing (KNL)} & \textbf{Broadwell}\\ 
    \hline
    \textbf{Model} & Intel Xeon Phi 3120P & Intel Xeon Phi 7250 & Intel Xeon E5-2699 v4 \\ 
    \textbf{Microarchitecture} & Intel Many Integrated Core & Intel Many Integrated Core & Intel Broadwell \\ 
    \textbf{Clock frequency} & 1.10 GHz & 1.40 GHz & 2.20 GHz \\
    \textbf{L1 cache (D/I)} & \unit{32}{\kibi\byte}/\unit{32}{\kibi\byte} & \unit{32}{\kibi\byte}/\unit{32}{\kibi\byte} & \unit{32}{\kibi\byte}/\unit{32}{\kibi\byte} \\
    \textbf{L2 cache} & \unit{30}{\mebi\byte} & \unit{34}{\mebi\byte} & \unit{256}{\kibi\byte} \\ 
    \textbf{L3 cache} & - & - & \unit{55}{\mebi\byte} \\
    \textbf{Cores/Threads} & 57/228 & 68/272 & 22/44 \\
    \textbf{Sockets} & 1 & 1 & 1 \\
    \textbf{STREAM triad main/llc~\cite{mccalpin1995stream}} & \unit{128/140}{\giga\byte\per\second} & \unit{395/570}{\giga\byte\per\second} & \unit{60/200}{\giga\byte\per\second} \\
    \hline
  \end{tabular}
  \vspace{-6mm}
\end{table}

\subsection{Feuture-Guided Classifier Accuracy}
\label{sec:accuracy}
First, we evaluate our feature-guided classifier in terms of accuracy,
assuming the labels generated by the profile-guided classifier are
correct. We estimate how accurately our models perform using
Leave-One-Out cross validation. According to this methodology, for a
training set of $k$ matrices ($k = 210$), $k$ experiments are
performed. For each experiment $k-1$ matrices are used for training
and one for testing. The computed performance metric per experiment is
the Exact Match Ratio, which is defined as the percentage of samples
for which the predicted set of classes is fully correct, i.e. it
exactly matches the set of labels produced by our profile-guided
classifier. However, since we apply at least one optimization per
matrix we can tolerate predictions that are partially correct, thus,
we also define a performance metric, denoted by \emph{Partial Match
  Ratio}, that considers a prediction to be correct if it contains at
least one correct class. The final accuracy score reported by cross
validation is the average of the values computed in the loop of $k$
experiments.


Table~\ref{table:classifiers:phi} reports the most accurate
feature-guided classifiers on KNC. Due to space limitations, we do not
present the classifiers of the other platforms, however, they employ a
similar set of features and similar accuracies. We report classifiers
with increasing time complexity in the feature extraction phase along
with their achieved accuracy score based on both the Exact and Partial
Match Ratio performance metrics. The selection of features for the
classifiers has been a result of exhaustive search.

\begin{table}
  \caption{Feature-guided Decision Tree classifiers on KNC}
  \label{table:classifiers:phi}
  \centering
  \begin{tabular}{|C{3.0cm}|C{1.4cm}|C{1.2cm}|C{1.3cm}|}
    \hline
    \textbf{Features} & \textbf{Complexity} & \textbf{Accuracy Exact (\%)} & \textbf{Accuracy Partial (\%)} \\
    \hline
    $nnz_{\{min,max,sd\}}$ $bw_{avg}$ $dispersion_{\{avg,sd\}}$ & O(N) & 80 & 95\\
    \hline
    $size, bw_{\{avg,sd\}}$ $nnz_{\{min,max,avg,sd\}}$ $misses_{avg}, dispersion_{sd}$ & O($NNZ$) & 84 & 100 \\
    \hline
  \end{tabular}
  \vspace{-6mm}
\end{table}

\subsection{Performance of selected optimizations}
\label{sec:eval:performance}
Contrary to standard classification problems, our key measure of
success is not classifier accuracy---rather, we aim to maximize the
average performance improvement of our optimizer over the baseline
implementation, as well as the CSR \texttt{mkl\_dcsrmv()} kernel
available in the Intel MKL library. In this way, we designate the
value of optimization tuning for SpMV. We also evaluate the recently
introduced Inspector-Executor CSR \texttt{mkl\_sparse\_d\_mv()} kernel
of Intel MKL.

\begin{figure*}[t]
  \centering
  \subfloat[KNC]{\includegraphics[width=0.98\textwidth]{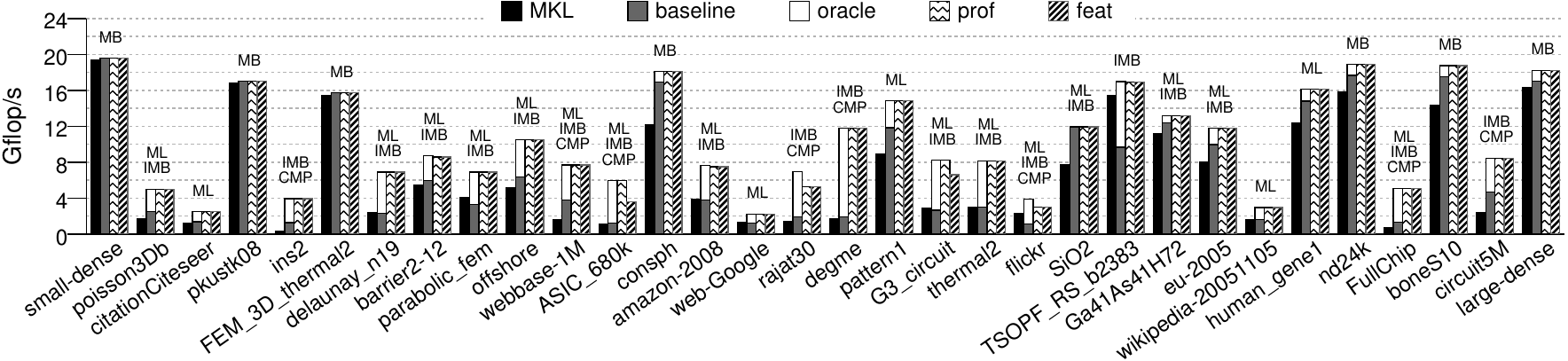}}\\
  \vspace{-2mm}
  \subfloat[KNL]{\includegraphics[width=0.98\textwidth]{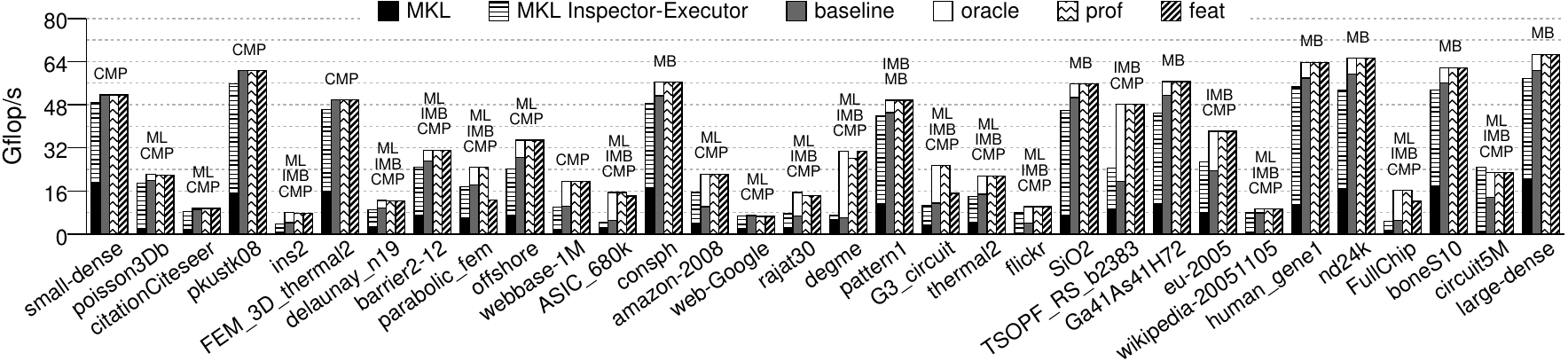}}\\
  \vspace{-2mm}
  \subfloat[Broadwell]{\includegraphics[width=0.98\textwidth]{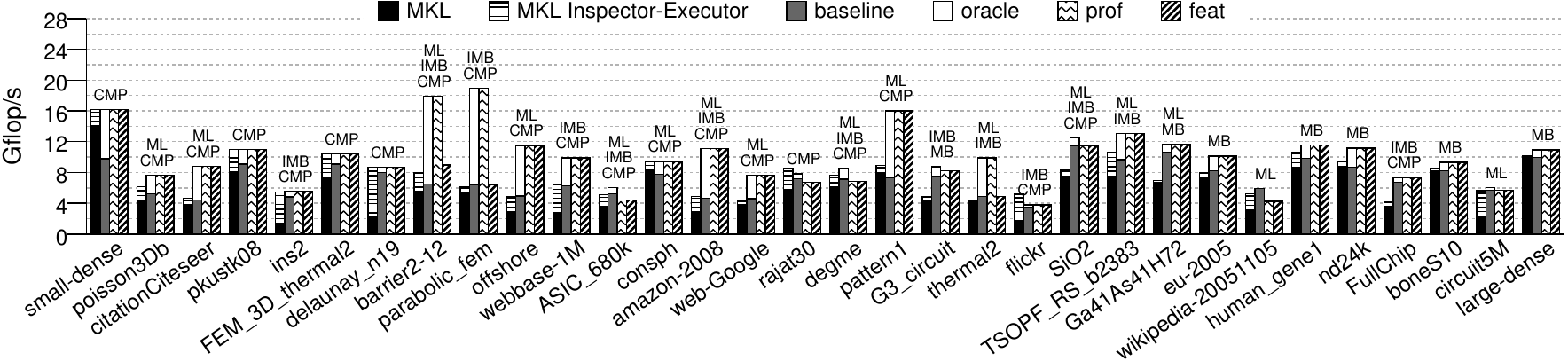}}
  \caption{SpMV performance landscape on each experimental
    platform. \emph{MKL} refers to the CSR SpMV kernel available in
    the Intel MKL library, \emph{MKL Inspector-Executor} refers to the
    autotuned CSR SpMV kernel available in the Intel MKL library,
    \emph{baseline} refers to our baseline CSR SpMV (see text in
    Section~\ref{sec:evaluation:setup}), \emph{feat} to the
    feature-guided optimizer, \emph{prof} to the profile-guided
    optimizer, and \emph{oracle} to the perfect optimizer. Note that
    MKL Inspector-Executor is not available on KNC.}
  \label{fig:performance}
  \vspace{-3mm}
\end{figure*}

Fig.~\ref{fig:performance} presents SpMV performance achieved by MKL
CSR, MKL Inspector-Executor CSR, our baseline CSR implementation, our
profiling- and feature-guided classifiers, as well as the
\emph{oracle}---the perfect optimizer that always selects the best
optimization available---on all experimental platforms for a
representative set of matrices. The classes of each matrix according
to the profile-guided classifier are also reported. We note that we
used the feature-guided classifier with the highest accuracy reported
in Table~\ref{table:classifiers:phi}.

On KNC and KNL, our initial observation concerns the diversity of
performance bottlenecks detected by our profile-guided
classifier. This was expected based on our bound and bottleneck
analysis in Section~\ref{sec:methodology:bounds}, and is mainly due to
the architectural characteristics of the Xeon Phi processors. Both KNC
and KNL contain a large number of cores, a fact that favors workload
imbalance, and a very expensive (an order of magnitude higher compared
to multi-cores) cache miss latency, which further exposes irregularity
in the accesses to the right-hand side vector. Thus, there are many
matrices that fall out of the standard MB class. Our classifiers
manage to successfully capture this trend and select proper
optimizations due to their matrix- and architecture-awareness. Notice
that the largest speedups for our classifiers are achieved by
combining optimizations from different classes.

Specifically on KNC, \emph{rajat30} and \emph{flickr} are the only
matrices where our classifiers do not detect the optimal
optimization. In case of \emph{rajat30}, which is characterized with a
high concentration of nonzero elements in a few dense rows, our
classifiers do not recognize it as a candidate of the ML class and,
consequently, do not apply software prefetching which offers the
additional performance boost. By further exploring, we discovered that
the benchmark that exposes irregularity for the profile-guided
classifier (see Section~\ref{sec:methodology:bounds}), can actually
detect the irregularity in this matrix by looking at it in partitions,
instead of looking at it as a whole. We intend to extend our
classification approach to incorporate this idea in future work. In
case of \emph{flickr}, the combination of \emph{auto} scheduling,
prefetching and vectorization did not pay off, with the best
performance being achieved by simply applying prefetching. In total,
the profile- and feature-guided classifiers achieve an impressive
average $2.72\times$ and $2.63\times$ speedup over MKL CSR
respectively.

On KNL, the profile- and feature-guided classifiers achieve an average
$6.73\times$ and $6.48\times$ speedup over MKL CSR respectively. The
MKL Inspector-Executor also significantly improves over MKL CSR by
$4.89\times$ on average, but our classifiers manage to further improve
performance in most cases. The largest speedups over the
Inspector-Executor occur for matrices with imbalanced execution, that
is either due to highly uneven row lengths (\{IMB, CMP\} set of
classes), e.g., \emph{ASIC\_680k, rajat30, degme}, or computational
unevenness (\{ML, IMB\} set of classes), e.g., \emph{parabolic\_fem,
  thermal2}. Also notice how some matrices present different or
additional bottlenecks compared to KNC. For instance,
\emph{human\_gene1} is latency bound (ML class) on KNC, while on KNL
it is bandwidth bound (MB class).

Finally, on Broadwell the profile- and feature-guided classifiers
achieve an average $2.02\times$ and $1.86\times$ speedup over MKL CSR
respectively. Speedups occur mostly for matrices that entirely or
partially fit in the system's cache hierarchy and include the ML or
IMB classes. The MKL Inspector-Executor also achieves an average
$1.49\times$ over MKL CSR, but our classifiers manage to further
improve performance in most cases.



\subsection{Runtime Overhead}
\label{sec:eval:overhead}
The work-flow of the proposed optimization process comprises two basic
steps: bottleneck detection through classification and generation of
optimized code based on the classification results. Concerning the
first step, our profile-guided classifier runs a number of
micro-benchmarks on the input matrix and applies the empirically-tuned
classification algorithm (see
Section~\ref{sec:methodology:classifiers:profiling}), while the
feature-guided classifier extracts matrix properties and applies a
pre-trained classifier (see
Section~\ref{sec:methodology:classifiers:feature}). Runtime code
generation is performed Just-In-Time (JIT). Depending on the selected
optimizations, e.g. compression, an intermediate preprocessing step
may be required.

In order to provide a meaningful illustration of how lightweight the
proposed optimizers are, we are going to evaluate them in the context
of iterative methods for the solution of large sparse linear systems,
e.g., variations of the Conjugate Gradient (CG) and the Generalized
Minimal Residual (GMRES) methods. Such solvers repeatedly call SpMV
and usually require hundreds to thousands of iterations to converge,
making it possible to amortize the overhead of the optimization
process, e.g., optimization selection, format conversion, runtime code
generation, parameter tuning etc. However, it is quite common in
real-life applications to run preconditioned versions of these methods
to accelerate convergence. In this case, the number of iterations may
be significantly smaller, ranging from dozens to hundreds, thus,
limiting the online overhead that can be tolerated~\cite{li2013gpu}.

In Table~\ref{table:overhead} we present the minimum number of
iterations required for our optimizers, as well as the MKL CSR
Inspector-Executor and two trivial optimizers---one that runs all
single optimizations (total of 5 in our case) and one that also
includes combinations of 2 (total of 15 in our case) and selects the
best for each matrix---to be beneficial over CSR MKL in the context of
an iterative solver. Specifically, if $t_{solver}$ and $t_{solver}'$
are the total execution times of the solver using a baseline and an
optimized SpMV implementation respectively, and $t_{pre}$ the
optimizer overhead, then the number of solver iterations required to
amortize that cost can be estimated as follows:
\begin{IEEEeqnarray}{rCl}
    t_{solver}' + t_{pre} && \ll t_{solver} \Leftrightarrow \nonumber \\
    N_{iters}\cdot t_{iter}' + t_{pre} && \ll N_{iters}\cdot t_{iter} \Leftrightarrow \nonumber \\
   N_{iters}\cdot (t_{other} + t_{spmv}') + t_{pre} && \ll N_{iters}\cdot
   (t_{other} + t_{spmv}) \Leftrightarrow \nonumber \\
   N_{iters}\cdot t_{spmv}' + t_{pre} && \ll N_{iters}\cdot t_{spmv} \Leftrightarrow \nonumber \\
   N_{iters} && \gg \frac{t_{pre}}{t_{spmv} - t_{spmv}'} \nonumber 
   \label{eq:2}
\end{IEEEeqnarray}
\noindent
where $N_{iters}$ is the number of iterations required for the solver
to converge, and $t_{spmv}$ and $t_{spmv}'$ the execution time of SpMV
before and after optimization. For simplicity, we have assumed that
the performance of other computations ($t_{other}$) in the solver are
not affected by changes in the SpMV implementation. By setting $t_{spmv} = t_{MKL}$ an $t_{spmv}' = t_{optimizer}$ we get:
\begin{IEEEeqnarray*}{rCl}
   N_{iters,min} && = \frac{t_{pre}}{t_{MKL} - t_{optimizer}}
   \label{eq:2}
\end{IEEEeqnarray*}
\noindent

In Table~\ref{table:overhead}, we report the worst and best observed
cases among our representative matrix suite, as well as the average
over all matrices. We run 64 SpMV iterations to get valid timing
measurements. Note that the trivial optimizers incur the preprocessing
overheads of every optimization, while our optimizers only incur the
overheads of the selected optimizations based on their predictions. As
expected, both optimizers that leverage the profiling- and
feature-guided classifiers have a significant advantage over the
trivial optimizers. The same stands for the MKL
Inspector-Executor. However, our feature-guided optimizer requires by
far the smallest number of iterations to provide a speedup over CSR
MKL, making it the most lightweight approach.

\begin{table}
  \renewcommand{\arraystretch}{1.15}
  \caption{Minimum number of solver iterations required to amortize the autotuning runtime overhead of different optimizers on KNL}
  \label{table:overhead}
  \centering
  \begin{tabular}{|c|c|c|c|c|}
    \hline
    \textbf{Optimizer} & \textbf{$N_{iters,best}$} & \textbf{$N_{iters,avg}$} & \textbf{$N_{iters,worst}$} \\
    \hline
    trivial-single & 455 & 910 & 8016 \\
    trivial-combined & 1992 & 3782 & 37111 \\
    \hline
    profile-guided & 145 & 267 & 3145 \\
    feature-guided & 27 & 60 & 567 \\
    MKL Inspector-Executor & 28 & 336 & 1229 \\
    \hline
  \end{tabular}
  \vspace{-3mm}
\end{table}

\section{Related Work}
\label{sec:related}
Different sparse matrices have different sparsity patterns, and
different architectures have different strengths and weaknesses. In
order to achieve the best SpMV performance for the target sparse
matrix on the target platform, an autotuning approach has long been
considered to be beneficial. The first autotuning approaches
attempted to tune parameters of specific sparse matrix storage
formats. Towards this direction, the Optimized Sparse Kernel Interface
(OSKI) library~\cite{vuduc2005oski} was developed as a collection of
high performance sparse matrix operation primitives on single core
processors. It relies on the SPARSITY framework~\cite{im2004sparsity}
to tune the SpMV kernel, by applying multiple optimizations, including
register blocking and cache blocking. Autotuning has also been used to
find the best block and slice sizes of the input sparse matrix on
modern CMPs and GPUs~\cite{choi2010model}.

There have been some research efforts closer to our work. The clSpMV
framework~\cite{su2012clspmv} is the first framework that analyzes the
input sparse matrix at runtime, and recommends the best representation
of the given sparse matrix, but it is restricted to GPU
platforms. Towards the same direction, the authors
in~\cite{guo2014performance} present an analytical and profiling-based
performance modeling to predict the execution time of SpMV on GPUs
using different sparse matrix storage formats, in order the select the
most efficient format for the target matrix. For each format under
consideration, they establish a relationship between the number of
nonzero elements per row in the matrix and the execution time of SpMV
using that format, thus encapsulating to some degree the structure of
the matrix in their methodology. Similarly,
in~\cite{li2015performance}, the authors propose a probabilistic model
to estimate the execution time of SpMV on GPUs for different sparse
matrix formats. They define a probability mass function to analyze the
sparsity pattern of the target matrix and use it to estimate the
compression efficiency of every format they examine. Combined with the
hardware parameters of the GPU, they predict the performance of SpMV
for every format. Since compression efficiency is the determinant
factor in this approach, it is mainly targeted for memory bandwidth
bound matrices. Closer to our approach is the SMAT autotuning
framework~\cite{li2013smat}. This framework selects the most efficient
format for the target matrix using feature parameters of the sparse
matrix. It treats the format selection process as a classification
problem, with each format under consideration representing a class,
and leverages a data mining approach to generate a decision tree to
perform the classification, based on the extracted feature parameters
of the matrix. The distinguishing advantage of our optimization
selection methodology over the aforementioned approaches, is that it
decouples the decision making from specific optimizations, by
predicting the major performance bottleneck of SpMV instead of SpMV
execution time using a specific optimization. Thus, in contrary to the
above frameworks, where incorporating a new optimization requires
either retraining a model or defining a new one, our decision-making
approach allows an autotuning framework to be easily extended, simply
by assigning the new optimization to one of the classes.


\section{Conclusions}
\label{sec:conclusion}
In this paper, we show how a matrix- and architecture-adaptive
optimizer can provide significant speedups for SpMV. The proposed
optimizer treats SpMV optimization as a classification problem. It
classifies the input matrix based on its performance bottlenecks and
applies suitable optimizations to tackle them. We describe two
classifiers: a) a profile-guided classifier that relies on online
profiling to perform the decision making, and b) a classifier trained
offline with machine learning techniques that only extracts structural
properties of the matrix on-the-fly. All aspects of the optimization
process focus on minimizing online overheads in order to be applicable
in real-life scenarios. Experimental evaluation on a representative
matrix suite on one multi- and two many-core platforms shows that our
methodology is very promising, achieving significant speedups over the
Intel MKL library with an online overhead that can be easily
amortized.

\section*{Acknowledgement}
This project has received funding from the European Union's Horizon 2020 research and innovation programme under grant agreement No 732204 (Bonseyes). This work was supported by the Swiss State Secretariat for Education, Research and Innovation (SERI) under contract number 16.0159. The opinions expressed and arguments employed herein do not necessarily reflect the official views of these funding bodies.



\bibliographystyle{IEEEtran}
\bibliography{IEEEabrv,paper}

\begin{thebibliography}{10}
\providecommand{\url}[1]{#1}
\csname url@samestyle\endcsname
\providecommand{\newblock}{\relax}
\providecommand{\bibinfo}[2]{#2}
\providecommand{\BIBentrySTDinterwordspacing}{\spaceskip=0pt\relax}
\providecommand{\BIBentryALTinterwordstretchfactor}{4}
\providecommand{\BIBentryALTinterwordspacing}{\spaceskip=\fontdimen2\font plus
\BIBentryALTinterwordstretchfactor\fontdimen3\font minus
  \fontdimen4\font\relax}
\providecommand{\BIBforeignlanguage}[2]{{%
\expandafter\ifx\csname l@#1\endcsname\relax
\typeout{** WARNING: IEEEtran.bst: No hyphenation pattern has been}%
\typeout{** loaded for the language `#1'. Using the pattern for}%
\typeout{** the default language instead.}%
\else
\language=\csname l@#1\endcsname
\fi
#2}}
\providecommand{\BIBdecl}{\relax}
\BIBdecl

\bibitem{toledo1997improving}
S.~Toledo, ``Improving the memory-system performance of sparse-matrix vector
  multiplication,'' \emph{IBM Journal of research and development}, vol.~41,
  no.~6, pp. 711--725, 1997.

\bibitem{pinar1999improving}
A.~Pinar and M.~T. Heath, ``Improving performance of sparse matrix-vector
  multiplication,'' in \emph{Proceedings of the 1999 ACM/IEEE conference on
  Supercomputing}.\hskip 1em plus 0.5em minus 0.4em\relax ACM, 1999, p.~30.

\bibitem{pichel2004improving}
J.~C. Pichel, D.~B. Heras, J.~C. Cabaleiro, and F.~F. Rivera, ``Improving the
  locality of the sparse matrix-vector product on shared memory
  multiprocessors,'' in \emph{Parallel, Distributed and Network-Based
  Processing, 2004. Proceedings. 12th Euromicro Conference on}.\hskip 1em plus
  0.5em minus 0.4em\relax IEEE, 2004, pp. 66--71.

\bibitem{vuduc2005fast}
R.~W. Vuduc and H.-J. Moon, ``Fast sparse matrix-vector multiplication by
  exploiting variable block structure,'' in \emph{High Performance Computing
  and Communications}.\hskip 1em plus 0.5em minus 0.4em\relax Springer, 2005,
  pp. 807--816.

\bibitem{willcock2006accelerating}
J.~Willcock and A.~Lumsdaine, ``Accelerating sparse matrix computations via
  data compression,'' in \emph{Proceedings of the 20th annual international
  conference on Supercomputing}.\hskip 1em plus 0.5em minus 0.4em\relax ACM,
  2006, pp. 307--316.

\bibitem{kourtis2008optimizing}
K.~Kourtis, G.~Goumas, and N.~Koziris, ``Optimizing sparse matrix-vector
  multiplication using index and value compression,'' in \emph{Proceedings of
  the 5th conference on Computing frontiers}.\hskip 1em plus 0.5em minus
  0.4em\relax ACM, 2008, pp. 87--96.

\bibitem{williams2009optimization}
S.~Williams, L.~Oliker, R.~Vuduc, J.~Shalf, K.~Yelick, and J.~Demmel,
  ``Optimization of sparse matrix--vector multiplication on emerging multicore
  platforms,'' \emph{Parallel Computing}, vol.~35, no.~3, pp. 178--194, 2009.

\bibitem{bulucc2009parallel}
A.~Bulu{\c{c}}, J.~T. Fineman, M.~Frigo, J.~R. Gilbert, and C.~E. Leiserson,
  ``Parallel sparse matrix-vector and matrix-transpose-vector multiplication
  using compressed sparse blocks,'' in \emph{Proceedings of the twenty-first
  annual symposium on Parallelism in algorithms and architectures}.\hskip 1em
  plus 0.5em minus 0.4em\relax ACM, 2009, pp. 233--244.

\bibitem{belgin2009pattern}
M.~Belgin, G.~Back, and C.~J. Ribbens, ``Pattern-based sparse matrix
  representation for memory-efficient smvm kernels,'' in \emph{Proceedings of
  the 23rd international conference on Supercomputing}.\hskip 1em plus 0.5em
  minus 0.4em\relax ACM, 2009, pp. 100--109.

\bibitem{kourtis2011csx}
K.~Kourtis, V.~Karakasis, G.~Goumas, and N.~Koziris, ``Csx: an extended
  compression format for spmv on shared memory systems,'' in \emph{ACM SIGPLAN
  Notices}, vol.~46, no.~8.\hskip 1em plus 0.5em minus 0.4em\relax ACM, 2011,
  pp. 247--256.

\bibitem{bulucc2011reduced}
A.~Bulu{\c{c}}, S.~Williams, L.~Oliker, and J.~Demmel, ``Reduced-bandwidth
  multithreaded algorithms for sparse matrix-vector multiplication,'' in
  \emph{Parallel \& Distributed Processing Symposium (IPDPS), 2011 IEEE
  International}.\hskip 1em plus 0.5em minus 0.4em\relax IEEE, 2011, pp.
  721--733.

\bibitem{kreutzer2014unified}
M.~Kreutzer, G.~Hager, G.~Wellein, H.~Fehske, and A.~R. Bishop, ``A unified
  sparse matrix data format for efficient general sparse matrix-vector
  multiplication on modern processors with wide simd units,'' \emph{SIAM
  Journal on Scientific Computing}, vol.~36, no.~5, pp. C401--C423, 2014.

\bibitem{li2013gpu}
R.~Li and Y.~Saad, ``Gpu-accelerated preconditioned iterative linear solvers,''
  \emph{The Journal of Supercomputing}, vol.~63, no.~2, pp. 443--466, 2013.

\bibitem{Greathouse:2014:ESM:2683593.2683678}
\BIBentryALTinterwordspacing
J.~L. Greathouse and M.~Daga, ``Efficient sparse matrix-vector multiplication
  on gpus using the csr storage format,'' in \emph{Proceedings of the
  International Conference for High Performance Computing, Networking, Storage
  and Analysis}, ser. SC '14.\hskip 1em plus 0.5em minus 0.4em\relax
  Piscataway, NJ, USA: IEEE Press, 2014, pp. 769--780. [Online]. Available:
  \url{https://doi.org/10.1109/SC.2014.68}
\BIBentrySTDinterwordspacing

\bibitem{Liu:2015:CES:2751205.2751209}
\BIBentryALTinterwordspacing
W.~Liu and B.~Vinter, ``Csr5: An efficient storage format for cross-platform
  sparse matrix-vector multiplication,'' in \emph{Proceedings of the 29th ACM
  on International Conference on Supercomputing}, ser. ICS '15.\hskip 1em plus
  0.5em minus 0.4em\relax New York, NY, USA: ACM, 2015, pp. 339--350. [Online].
  Available: \url{http://doi.acm.org/10.1145/2751205.2751209}
\BIBentrySTDinterwordspacing

\bibitem{saad1992numerical}
Y.~Saad, \emph{Numerical methods for large eigenvalue problems}.\hskip 1em plus
  0.5em minus 0.4em\relax SIAM, 1992, vol. 158.

\bibitem{williams2009roofline}
S.~Williams, A.~Waterman, and D.~Patterson, ``Roofline: an insightful visual
  performance model for multicore architectures,'' \emph{Communications of the
  ACM}, vol.~52, no.~4, pp. 65--76, 2009.

\bibitem{mellor2004optimizing}
J.~Mellor-Crummey and J.~Garvin, ``Optimizing sparse matrix--vector product
  computations using unroll and jam,'' \emph{International Journal of High
  Performance Computing Applications}, vol.~18, no.~2, pp. 225--236, 2004.

\bibitem{tang2015optimizing}
W.~T. Tang, R.~Zhao, M.~Lu, Y.~Liang, H.~P. Huyng, X.~Li, and R.~S.~M. Goh,
  ``Optimizing and auto-tuning scale-free sparse matrix-vector multiplication
  on intel xeon phi,'' in \emph{Code Generation and Optimization (CGO), 2015
  IEEE/ACM International Symposium on}.\hskip 1em plus 0.5em minus 0.4em\relax
  IEEE, 2015, pp. 136--145.

\bibitem{lazowska1984quantitative}
E.~D. Lazowska, J.~Zahorjan, G.~S. Graham, and K.~C. Sevcik, \emph{Quantitative
  system performance: computer system analysis using queueing network
  models}.\hskip 1em plus 0.5em minus 0.4em\relax Prentice-Hall, Inc., 1984.

\bibitem{davis2011university}
T.~A. Davis and Y.~Hu, ``The university of florida sparse matrix collection,''
  \emph{ACM Transactions on Mathematical Software (TOMS)}, vol.~38, no.~1,
  p.~1, 2011.

\bibitem{scikit-learn}
F.~Pedregosa, G.~Varoquaux, A.~Gramfort, V.~Michel, B.~Thirion, O.~Grisel,
  M.~Blondel, P.~Prettenhofer, R.~Weiss, V.~Dubourg, J.~Vanderplas, A.~Passos,
  D.~Cournapeau, M.~Brucher, M.~Perrot, and E.~Duchesnay, ``Scikit-learn:
  Machine learning in {P}ython,'' \emph{Journal of Machine Learning Research},
  vol.~12, pp. 2825--2830, 2011.

\bibitem{pooch1973survey}
U.~W. Pooch and A.~Nieder, ``A survey of indexing techniques for sparse
  matrices,'' \emph{ACM Computing Surveys (CSUR)}, vol.~5, no.~2, pp. 109--133,
  1973.

\bibitem{dagum1998openmp}
L.~Dagum and R.~Enon, ``Openmp: an industry standard api for shared-memory
  programming,'' \emph{Computational Science \& Engineering, IEEE}, vol.~5,
  no.~1, pp. 46--55, 1998.

\bibitem{mccalpin1995stream}
J.~D. McCalpin, ``Stream: Sustainable memory bandwidth in high performance
  computers,'' 1995.

\bibitem{vuduc2005oski}
R.~Vuduc, J.~W. Demmel, and K.~A. Yelick, ``Oski: A library of automatically
  tuned sparse matrix kernels,'' in \emph{Journal of Physics: Conference
  Series}, vol.~16, no.~1.\hskip 1em plus 0.5em minus 0.4em\relax IOP
  Publishing, 2005, p. 521.

\bibitem{im2004sparsity}
E.-J. Im, K.~Yelick, and R.~Vuduc, ``Sparsity: Optimization framework for
  sparse matrix kernels,'' \emph{International Journal of High Performance
  Computing Applications}, vol.~18, no.~1, pp. 135--158, 2004.

\bibitem{choi2010model}
J.~W. Choi, A.~Singh, and R.~W. Vuduc, ``Model-driven autotuning of sparse
  matrix-vector multiply on gpus,'' in \emph{ACM Sigplan Notices}, vol.~45,
  no.~5.\hskip 1em plus 0.5em minus 0.4em\relax ACM, 2010, pp. 115--126.

\bibitem{su2012clspmv}
B.-Y. Su and K.~Keutzer, ``clspmv: A cross-platform opencl spmv framework on
  gpus,'' in \emph{Proceedings of the 26th ACM international conference on
  Supercomputing}.\hskip 1em plus 0.5em minus 0.4em\relax ACM, 2012, pp.
  353--364.

\bibitem{guo2014performance}
P.~Guo, L.~Wang, and P.~Chen, ``A performance modeling and optimization
  analysis tool for sparse matrix-vector multiplication on gpus,''
  \emph{Parallel and Distributed Systems, IEEE Transactions on}, vol.~25,
  no.~5, pp. 1112--1123, 2014.

\bibitem{li2015performance}
K.~Li, W.~Yang, and K.~Li, ``Performance analysis and optimization for spmv on
  gpu using probabilistic modeling,'' \emph{Parallel and Distributed Systems,
  IEEE Transactions on}, vol.~26, no.~1, pp. 196--205, 2015.

\bibitem{li2013smat}
J.~Li, G.~Tan, M.~Chen, and N.~Sun, ``Smat: an input adaptive auto-tuner for
  sparse matrix-vector multiplication,'' in \emph{ACM SIGPLAN Notices},
  vol.~48, no.~6.\hskip 1em plus 0.5em minus 0.4em\relax ACM, 2013, pp.
  117--126.

\end{thebibliography}
%



\end{document}